\documentstyle[12pt]{article}
\begin{document}
\font\rm=cmr12
\font\tenrm=cmr10
\font\tensl=cmsl10
\font\tenbf=cmbx10
\rightline{}
\rightline{}
\rightline{}
\rightline{INFN-ISS 94/7}
\rightline{July 94}

\vspace{4cm}
\begin{center}
{}~~~~~
\tenrm
{\tenbf   INVESTIGATION OF THE NEUTRON FORM FACTORS BY INCLUSIVE\\
QUASI-ELASTIC SCATTERING OF POLARIZED ELECTRONS\\OFF POLARIZED
$^{3}$He: A THEORETICAL OVERVIEW  }
\\[0.8cm]
C. CIOFI degli ATTI\\
{\tensl Department of Physics, University of
Perugia, and INFN, Sezione di Perugia,\\
Via A. Pascoli, I-06100 Perugia, Italy}\\[0.5cm]
E. PACE\\
{\tensl Dipartimento di Fisica, Universit\`a di Roma "Tor Vergata",
and INFN, \\
Sezione Tor Vergata, Via della Ricerca Scientifica, I-00133 Roma, Italy}
\\[0.3cm]
and
\\[0.3cm]
 G. SALM\`E
\\
{\tensl INFN, Sezione Sanit\`a, Viale Regina Elena 299, I-00161
Roma, Italy}
\\[.8cm]
ABSTRACT\\[0.5cm]
\end{center}
{\tenrm{ The theory of quasi-elastic inclusive scattering of polarized leptons
off polarized $^3$He is critically reviewed and the origin of different
expressions for the polarized nuclear response function appearing in the
literature is explained. The sensitivity of the longitudinal asymmetry upon
the
neutron form factors is thoroughly investigated and the role played by the
polarization angle for minimizing the proton contribution is illustrated.}}

\bigskip
{\bf{PACS:}} 25.30.-c,24.70.+s,25.10.+s,29.25.pg

\bigskip

Phys. Rev. C   in press
\newpage
\bigskip
\noindent{\bf 1. Introduction}
\bigskip

\indent  The scattering of polarized electrons by
polarized targets represents a valuable tool for investigating
nucleon and nuclear properties in great detail \cite{1.}. In particular
quasi-elastic (qe) inclusive experiments involving polarized $^3$He  are
viewed as possible source of information on the neutron form factors
\cite{2.,3.,gao}. As a matter of fact, if a naive model of $^3$He, including
only the main spatially symmetric S component of the three-body wave function
with the two protons with opposite spins, is considered, a polarized $^{3}$He
can
be regarded to a large extent as an effective neutron target. However a
proton contribution, arises from the S' and D-waves of the three body wave
function. Such a contribution has been investigated in Ref.\cite{4.} within the
closure approximation, i.e., by describing nuclear effects through
spin-dependent
momentum distributions. Adopting the general formalism of Ref.\cite{4.}, the
effects
of nucleon binding have been analysed in Ref.\cite{5.}, where the concept of
the
spin dependent spectral function has been introduced and applied to the
calculation of the $^3$He asymmetry.  An analysis of the
asymmetry in plane wave impulse
approximation (PWIA) has been performed also in Ref.\cite{6.}, where, besides
studying the effects of binding: i) a new
expression for the polarized nuclear structure functions has been obtained and
the formalism of Ref.\cite{5.} has been shown to suffer from severe
inconsistencies, and ii) doubts have been raised as to the possibility of
obtaining reliable information on the neutron form factors by the measurement
of
the inclusive asymmetry, due to the large proton contribution.

\indent In view of the relevance of this two points it is our aim: i) to
elucidate in detail the origin of the
differences between the expression of the polarized structure functions used in
Refs.\cite{4.,5.} and the one obtained in Ref.\cite{6.}  by presenting
a comprehensive derivation of the inclusive cross section in PWIA, and ii) to
show that the proton contribution  depends upon the kinematics in such a way
that  one can choose a proper kinematics in order to make  the qe asymmetry
very
sensitive to  neutron properties, including the neutron electric form factor.

\indent The paper is organized as follows: in Sect. 2 the antisymmetric part
of the hadronic tensor will be analyzed and the different methods for
obtaining the nuclear polarized structure functions will be discussed; in Sect.
3 the nuclear polarized structure functions will be obtained within the
 PWIA by using the nucleon spin-dependent spectral function; in Sect. 4 the
comparison with the experimental asymmetries of Ref.\cite{2.,3.} will be
discussed; in Sect. 5 a proposal for minimizing the proton contribution will be
illustrated and in Sect. 6 conclusions will be drawn. \bigskip

\pagestyle{plain}
\noindent{\bf 2. The hadronic tensor and the inclusive cross section} \bigskip

\indent In what follows we will consider the inclusive cross section describing
the scattering of a longitudinally polarized lepton of helicity $h~=~\pm1$ by a
polarized hadron of spin J = 1/2; in one photon exchange approximation
one gets \cite{1.}

\begin{eqnarray}
\frac{d^2\sigma(h)}{d\Omega_2 d\nu}~\equiv~
\sigma_2\left(\nu,Q^2,\vec{S}_A,h\right)& =& \frac{4 \alpha^2}{Q^4}\;
\frac{\epsilon_2}{\epsilon_1}\;m^2\;
L^{\mu\nu}W_{\mu\nu}\;=\nonumber\\
&=&\; \frac{4 \alpha^2}{Q^4}\;
\frac{\epsilon_2}{\epsilon_1}\;m^2\;
\left[L^{\mu\nu}_{s}W_{\mu\nu}^{s}+L^{\mu\nu}_{a}
W_{\mu\nu}^{a}\right]
\label{eq1}
\medskip
\end{eqnarray}
where the  symmetric ($s$) and antisymmetric ($a$) leptonic tensors
are

\begin{eqnarray}
 L^{\mu\nu}_{s} &=&\; -(g^{\mu\nu}
+\frac{q^\mu\;q^\nu}{Q^2})\;\frac{Q^2}{4m^2} +~{1 \over m^2}
(k^\mu_1-\frac{q^\mu}{2})\;(k^\nu_1-\frac{q^\nu}{2})
\label{2.1}\\
 L^{\mu\nu}_{a}
&=&\;i~h~\epsilon^{\mu\nu\rho\sigma}\frac{q_\rho\;k_{1\sigma}}{2m^2}
\label{eq2.2} \medskip \end{eqnarray}
and the corresponding  hadronic  tensors are

\begin{eqnarray}
 W_{\mu\nu}^{s} &=&\; -(g_{\mu\nu} +\frac{q_\mu\;q_\nu}{Q^2})\;W^A_1 +
(P_{A\mu}+\frac{P_A\cdot q}{Q^2}q_\mu)\;(P_{A\nu}+\frac{P_A\cdot
q}{Q^2}q_\nu)~\frac{W^A_2}{M^2_A}
\label{3.1} \\
 W_{\mu\nu}^{a}
&=&\;i\epsilon_{\mu\nu\rho\sigma}q^\rho\;V^\sigma
 \label{eq3.2}
\medskip \end{eqnarray}
The pseudovector  $V^\sigma$ appearing in Eq.(\ref{eq3.2}) can be expressed as
follows \begin{equation}
V^\sigma\equiv~S^\sigma_A~\frac{G^A_1}{M_A} +
(P_A\cdot q\;S^\sigma_A - S_A\cdot q\;P^\sigma_A)
{}~\frac{G^A_2}{M^3_A}\label{eq3.3}
\medskip
\end{equation}
In the above equations, the index A
 denotes the number of nucleons  composing
the target, $k^\mu_{1(2)} \equiv (\epsilon_{1(2)},\vec{k}_{1(2)})$
and $P_A^\mu  \equiv (M_A,0)$ are electron and target four-momenta,
$~q^\mu \equiv (\nu,\vec{q})$ is the four-momentum transfer, $Q^2 =
-q^2=|\vec{q}|^2-\nu^2$,  $~g_{\mu\nu}$  is the symmetric metric tensor,
$\epsilon_{\mu\nu\rho\sigma}$ the fully antisymmetric tensor, $~S_A^\mu$ the
polarization four-vector (in the nucleus rest frame $S_A^\mu \equiv
(0,\vec{S}_A)$) and $W^A_{1(2)}$ and $G^A_{1(2)}$ are the nuclear unpolarized
and
polarized structure functions, respectively.

\indent In  polarized scattering, both the symmetric
($W^{s}_{\mu\nu}$) and the antisymmetric ($W^{a}_{\mu\nu}$) part of the
hadronic tensor are involved, but in what follows we will focus on the
antisymmetric one, since it contains the relevant physical quantities
we will investigate. To this end the following comments are in order:
\begin{itemize}
\item[i)] to obtain the  general form of $W^{a}_{\mu\nu}$ one follows a
 procedure analogous to the one adopted to obtain the symmetric tensor
$W^{s}_{\mu\nu}$, namely one imposes Lorentz, gauge, parity
and time reversal invariances
on the weighted sum of all the available {\em antisymmetric tensors};
\item[ii)] terms
proportional to $q^\sigma$,  which in principle could appear in
the definition of $V^\sigma$, were not included in Eq. (\ref{eq3.3})
since the antisymmetric tensor $\epsilon_{\mu\nu\alpha\beta}$ in
Eq. (\ref{eq3.2}) cancels out the contribution to $W^{a}_{\mu\nu}$ arising from
any term of this
kind ( $\epsilon_{\mu\nu\alpha\beta} q^\alpha q^\beta~=~0$).
\end{itemize} Let us stress that, according to remark ii), only the component
of $V^\sigma$ orthogonal to $q^\sigma$ can be determined by the knowledge of
the tensor $W^{a}_{\mu\nu}$, and  therefore only such a component represents a
physically relevant quantity, while the one  parallel to
$q^\sigma$ is completely undetermined. Indeed, by  "inverting" Eq.(\ref{eq3.2})
one obtains

\begin{equation}
\tilde{V}^\sigma~\equiv~V^\sigma~+ {q^\sigma\over Q^2}~(V\cdot q)~=~i {1\over
{2~Q^2}}~\epsilon^{\sigma\alpha\mu\nu}q_\alpha~W^{a}_{\mu\nu}
\label{eq4}\medskip
\end{equation}
where the four-vector $\tilde{V}^\sigma$ results to be orthogonal to
$q^\sigma$, i.e. $\tilde{V}\cdot q~=~0$. The relevance of this last comment
will
be clear later on, when the  method for obtaining  the polarized structure
functions within the PWIA
will be discussed in detail.

\indent
In total analogy with the case of the unpolarized
structure functions $W^A_{1(2)}$, which are determined by the symmetric part of
the hadronic tensor, the polarized structure functions  $G_{1(2)}^A$  are
obtained by expressing them in terms of the components of  $W^{a}_{\mu\nu}$, or
equivalently,  they can be expressed in terms of the components of the
pseudovector $\tilde{V}^\sigma$.
 In the rest frame of the target,
   using
Eqs. (\ref{eq3.2}), (\ref{eq3.3}) and (\ref{eq4}) and choosing the
z-axis  along the momentum transfer  ($\hat{q}~\equiv~\hat{u}_z$), one gets the
following expressions for the polarized structure functions $G_1^A$ and
$G_2^A$
(cf. Refs.\cite{6.} and \cite{7.})

\begin{eqnarray}
{G_1^A\over M_A}&=&~-~{Q^2\over |\vec{q}|^2} \left({\tilde{V}_z\over S_{Az}}~-~
{\tilde{V}_\perp \over S_{A\perp}}\right) =
-{i\over |\vec{q}|} \left({Q^2\over
|\vec{q}|^2}~{W^{a}_{02}\over S_{Ax}}~+~{\nu \over |\vec{q}|} {W^{a}_{12}\over
S_{Az}}\right)    \label{5.1} \\
{G_2^A\over M_A^2}&=&{\tilde{V}_0 \over |\vec{q}|S_{Az}}~-~{\nu\over
|\vec{q}|^2}
 \left({\tilde{V}_z\over
S_{Az}}~-~ {\tilde{V}_\perp\over S_{A\perp}}\right)
=-{i\over
|\vec{q}|^2}~\left({\nu \over |\vec{q}|}{W^{a}_{02}\over S_{Ax}}~-~
{W^{a}_{12}\over S_{Az}}\right)  \label{5.2}
\medskip
\end{eqnarray}
where $S_{A\perp}=\sqrt{S_{Ax}^2+S_{Ay}^2}$ and
$\tilde{V}_{\perp}=(\vec{\tilde{V}} \cdot
\vec{S}_A-\tilde{V}_zS_{Az})/S_{A\perp}$.

\noindent  In line with
remark ii),   Eqs.(\ref{5.1}) and (\ref{5.2}) are not affected, because of
Eq.(\ref{eq3.2}),
 by any arbitrary term proportional to $q^\sigma$ which  could be added to
$V^\sigma$. In Ref.\cite{4.} a different
extraction method was proposed, namely $G_{1(2)}^A$ were obtained by expressing
them in terms of the components of the pseudovector ${V}^\sigma$.
{}From Eq.(\ref{eq3.3}) one has

\begin{eqnarray}
{G_1^A\over M_A}&=&-{(V\cdot q)\over{|\vec{q}|S_{Az}}}
\label{5.3} \\
{G_2^A\over M_A^2}&=&{V_0\over{|\vec{q}|S_{Az}}}
\label{5.4}
\medskip
\end{eqnarray}
Given the form (\ref{eq3.3}) for $V^\sigma$, Eqs. (\ref{5.1}) and (\ref{5.2})
are totally equivalent to Eqs. (\ref{5.3}) and (\ref{5.4}). However, such an
equivalence will break down if a term proportional to $q^\sigma$ is explicitely
added to the r.h.s. of Eq. (\ref{eq3.3}), since, as already noted,
Eqs. (\ref{5.1}) and (\ref{5.2}) will be unaffected by the added term, whereas
Eqs. (\ref{5.3}) and (\ref{5.4}) will be; therefore  $G_1^A$ and  $G_2^A$
obtained from  Eqs. (\ref{5.3}) and (\ref{5.4}) will be  incorrect in this
case.
This remark will be very relevant for the discussion of the evaluation of
$G_1^A$
and  $G_2^A$ within the PWIA, which will be presented in the next Section. To
sum
up, unlike the unpolarized case, two different procedures have been followed to
obtain the polarized structure functions $G_{1(2)}^A$; they lead to
Eqs. (\ref{5.1}) and (\ref{5.2}) and Eqs. (\ref{5.3}) and (\ref{5.4}),
respectively; however the latters are correct only in so far as $V^\sigma$ is a
linear combination of only $S^\sigma_A$ and  $P^\sigma_A$ and terms
proportional
to $q^\sigma$ are absent. We will call \cite{7.} the correct prescription
leading
to Eqs. (\ref{5.1}) and  (\ref{5.2}) prescription I (corresponding to the
prescription A of Ref. \cite{6.}) and the one leading to Eqs. (\ref{5.3}) and
(\ref{5.4}) prescription II (corresponding  to the prescription C of Ref.
\cite{6.} and originally proposed in Ref. \cite{4.}).

\bigskip

\noindent{\bf 3. The polarized structure functions in PWIA}
\bigskip

\indent The equations given in Sect.2 are general ones, relying only on the one
photon exchange approximation. When comparing with experimental data, one
has to adopt models for the nuclear structure functions. In particular  all
papers so far published \cite{4.,5.,6.,7.} use the PWIA. Within such an
approximation, one can obtain the following expression for the antisymmetric
hadronic tensor

\begin{equation}
{w}^{a}_{\mu\nu}~=~i\epsilon_{\mu\nu\alpha\beta}~q^\alpha~R^\beta
\label{eq6}
\end{equation}
where the four-pseudovector $R^\beta$ is given by

\begin{eqnarray}
R^\beta~=~\sum\limits_{N=p,n}\left[ {\tilde{G}^N_1(Q^2) \over
M}~\langle~S^\beta~\rangle_N + {\tilde{G}^N_2(Q^2) \over
M^3}~q_\alpha\left
(\langle~p^\alpha~S^\beta\rangle{_N}-\langle~p^\beta~S^\alpha\rangle{_N}\right)\right]
\label{7}
\end{eqnarray}
In Eq. (\ref{7}) $p^\alpha~\equiv~(\sqrt{M^2+{|\vec{p}|}^2},\vec p)$ is the
on-shell nucleon momentum, $\tilde{G}^{N}_{1}(Q^2)$ and
$\tilde{G}^{N}_{2}(Q^2)$ are the nucleon polarized form
factors, related to the nucleon Sachs form factors by the following
equations \cite{4.}

\begin{eqnarray}
\tilde{G}^{N}_{1}(Q^2)&=&-~{G_M^{N}\over
2}~{(G_E^{N}+\tau~G_M^{N})\over (1+\tau)} \label{7.1} \\
\tilde{G}^{N}_{2}(Q^2)&=&{G_M^{N}\over
4}~{(G_M^{N}-~G_E^{N})\over (1+\tau)}
\label{7.2}
\medskip
\end{eqnarray}
 with $\tau=Q^2/(4M^2)$, and the mean values
$\langle~S^\beta\rangle{_{N}}~$ and
$~\langle~p^\alpha~S^\beta\rangle{_{N}}$ read as follows

\begin{eqnarray}
\langle~[p^\alpha]~S^\beta\rangle{_{N}}~&=&~\int~dE~\int
d\vec{p}~{M^2\over
E_p~E_{p+q}}~[p^\alpha]~\sum\limits_{\ell=x,y,z}~f^{N}_{{\cal
M},\ell}(\vec{p},E)~S^\beta_\ell \nonumber \\
&~&\delta(\nu+M_A-\sqrt{(M~(A-1)+E_{f_{(A-1)}})^2+{|\vec{p}|}^2}-E_{p+q})
\label{eq7.3}
\medskip
\end{eqnarray}
In
Eq. (\ref{eq7.3})
$E_{p+q}=\sqrt{M^2+(\vec {p}+ \vec{q})^2}$, $E=E_{f_{(A-1)}}-E_A$ is the
nucleon
removal energy, $E_{f_{(A-1)}}$ and $E_A$ are the eigenvalues of the energy of
the spectator system and of the target nucleus, respectively, and
$S^\beta_\ell~\equiv~(\hat{u}_\ell\cdot \vec{p}/
M,\hat{u}_\ell+\vec{p}~\hat{u}_\ell\cdot \vec{p}/( M(E_p +M))~)$ is a
four-vector, which in the rest frame of the nucleon has
 the direction of the $\ell$-axis (  $\hat{u}_\ell$  is the
versor corresponding to the $\ell$-axis ($\ell~=~x,y,z$)).
Eq. (\ref{eq7.3}) is a generalization \cite{5.,7.} of the expressions of Ref.
\cite{4.} to the case where both the nucleon momentum and energy distributions
are considered.

\indent In
Eq. (\ref{eq7.3}) the three-dimensional pseudovector $\vec{f}^{N}_{\cal
M}(\vec{p},E)$ describes the nuclear structure and is  defined as follows

\begin{eqnarray}
\vec{f}^{N}_{\cal
{M}}(\vec{p},E)~=~{\hbox{\large{Tr}}}
 \left(~{\bf\hat{P}}^{N}_{\cal{M}}(\vec{p},E)
{}~\vec{\sigma} \right )
\medskip
\label{eq8}
\end{eqnarray}
where the 2x2 matrix ${\bf\hat{P}}^{N}_{\cal{M}}(\vec{p},E)$ is the spin
dependent spectral function of a nucleon inside a nucleus with  polarization
$\vec{S}_A$ oriented, in general, along a direction different from the z-axis,
and $\cal M$ {\em is the  component of the total angular momentum along
$\vec{S}_A$}. The elements of the matrix
${\bf\hat{P}}^{N}_{\cal{M}}(\vec{p},E)$
are  given by
\begin{eqnarray}
 P_{\sigma,
\sigma',\cal{M}}^{N} ({\vec{p},E})=\sum\nolimits\limits_{{f}_{(A-1)}}
{}~_{N}\langle{\vec{p},\sigma;\psi
}_{f_{(A-1)}} |{\psi }_{J\cal{M}}\rangle~\langle{\psi
}_{J\cal{M}}|{\psi }_{f_{(A-1)}};\vec{p},\sigma '\rangle _{N}~
\delta (E-{E}_{f_{(A-1)}}+{E}_{A})
\label{eq9}
\medskip
\end{eqnarray}
where $|{\psi
}_{J\cal{M}}\rangle$ is the ground state of the target nucleus polarized along
$\vec{S}_A$, $|{\psi }_{f_{(A-1)}}\rangle$  an eigenstate of the (A-1) nucleon
system interacting with the same two-body potential of the target nucleus,
$|\vec{p},\sigma\rangle_N$  the plane wave for the nucleon $N$ with the spin
along the z-axis equal to $\sigma$. In a more compact form, for $J=1/2$,
${\bf\hat{P}}^{N}_{\cal{M}}(\vec{p},E)$ is given by
\begin{eqnarray}
{\bf\hat{P}}^{N}_{\cal{M}}(\vec{p},E)={1\over
2}\left[B_{0,{\cal{M}}}^{N}(|\vec{p}|,E)~+~\vec{\sigma} \cdot \vec{f}^{N}_{\cal
{M}}(\vec{p},E)\right]
\label{eq9.1}
\medskip
\end{eqnarray}
where the function $B_{0,\cal{M}}^{N}(|\vec{p}|,E)$ is the trace of
${\bf\hat{P}}^{N}_{\cal{M}}(\vec{p},E)$ and yields the
usual unpolarized spectral function $P^{N}\!\left( |\vec{p}| , E\right)~$
\cite{8.}. It should be noticed that the matrix
${\bf\hat{P}}^{N}_{\cal{M}}(\vec{p},E)$ and the pseudovector
$\vec{f}^{N}_{\cal{M}}(\vec{p},E)$ depend on the direction of the polarization
vector $\vec{S}_A$. Since   $\vec{f}^{N}_{\cal{M}}(\vec{p},E)$ is a
pseudovector, it is a linear combination of the pseudovectors at our disposal,
viz. $\vec{S}_A$ and $\hat{p}~(\hat{p} \cdot \vec{S}_A)$, and therefore it can
be put in the following form, where any angular dependence is explicitely
given,

\begin{eqnarray}
\vec{f}^{N}_{\cal{M}}(\vec{p},E)~=~\vec{S}_A~B_{1,\cal{M}}^{N}(|\vec{p}|,E)~+~\hat{p}~(\hat{p}
\cdot \vec{S}_A)~ B_{2,\cal{M}}^{N}(|\vec{p}|,E)
\label{eq10}
\medskip
\end{eqnarray}
 The
functions $B_{0,\cal{M}}^{N}(|\vec{p}|,E)$,  $B_{1,\cal{M}}^{N}(|\vec{p}|,E)$
and
$B_{2,\cal{M}}^{N}(|\vec{p}|,E)$ satisfy the following relations
 \begin{eqnarray} B_{0,{1 \over 2}}^{N}(|\vec{p}|,E)&=&B_{0,-{1 \over
2}}^{N}(|\vec{p}|,E) \nonumber \\
B_{1,{1 \over 2}}^{N}(|\vec{p}|,E)&=&-B_{1,-{1 \over
2}}^{N}(|\vec{p}|,E)  \\
B_{2,{1 \over 2}}^{N}(|\vec{p}|,E)&=&-B_{2,-{1 \over
2}}^{N}(|\vec{p}|,E))
\nonumber
\medskip
\end{eqnarray}
The explicit expressions of the functions
$B^{N}_{0(1,2),\cal{M}}$ for a nucleus
with an arbitrary value of the total angular momentum are given in the
Appendix A in terms of the overlap integrals
{}~$_{N}\langle{\vec{\rho},\sigma;\psi }_{f_{(A-1)}} |{\psi
}_{J\cal{M}}\rangle$,
where $\rho$ is the Jacobi coordinate of the nucleon $N$ with respect to the
(A-1) system.

The spin dependent spectral function of $^3$He  has been first
obtained \cite{5.} from the overlap integrals, calculated
with a variational three body wave function corresponding to the Reid soft-core
interaction \cite{10.}. The same quantity has been calculated in Ref.
\cite{6.}, but
using a Faddeev wave function and the Paris potential \cite{11.}, obtaining a
similar $|\vec{p}|$ and $E$ dependence. In Fig. 1 (2), the unpolarized spectral
function  $P^{p(n)}\!\left(|\vec{p}| , E\right) \equiv B_{0,{1
\over 2}}^{p(n)}(|\vec{p}|,E)$ and the  functions $|B_{1,{1
\over 2}}^{p(n)}(|\vec{p}|,E)|~$ and $|B_{2,{1 \over 2}}^{p(n)}(|\vec{p}|,E)|$
are shown (in the case of the proton the curve corresponding to a spectator
deuteron is not presented).
 The relations between $B_{1,{\cal{M}}}^{N}$, $B_{2,{\cal{M}}}^{N}$ and
the quantities ${\hbox{\Large\it
P}}_{\parallel}^{N}\!\left(|\vec{p}|,E,\alpha\right)$
 and ${\hbox{\Large\it P}}_{\perp}^{N}\!\left( |\vec{p}|,E,\alpha
\right)$ used in our previous paper \cite{5.} can be found from
Eqs.(\ref{eq8}), (\ref{eq9}), (\ref{eq9.1}) and (\ref{eq10}) by assuming
$\vec{S}_A~\equiv\hat{q}\equiv \hat{u}_z$ and $\cal{M}$ = 1/2. Indeed, from the
z-component of $\vec{f}^{N}_{\cal{M}}(\vec{p},E)$ one has

 \begin{eqnarray}
{\hbox{\Large\it P}}_{\parallel}^{N}\!\left( |\vec{p}|,E,\alpha\right)
&=&{{P}_{{1
\over 2},{1 \over 2},{1 \over 2}}^{N}\left({\vec{p},E}\right)-{P}_{-{1 \over
2},-{1 \over 2},{1 \over 2}}^{N}\left(\vec{p},E\right)}\nonumber \\
&=&B^{N}_{1,{1 \over 2}} (|\vec{p}|,E)~+~B^{N}_{2,{1 \over 2}}
(|\vec{p}|,E)~cos^2\alpha \label{eq10.1}
\medskip
\end{eqnarray}
while from the other two components of $\vec{f}^{N}_{\cal{M}}(\vec{p},E)$ one
gets \begin{eqnarray}
{\hbox{\Large\it P}}_{\perp}^{N}\!\left( |\vec{p}|,E,\alpha\right)&=&2~{P}_{{1
\over 2},-{1 \over 2},{1
\over 2}}^{N}\left(\vec{p},E\right)~e^{i\phi}\nonumber \\
&=&B^{N}_{2,{1 \over 2}}(|\vec{p}|,E)~cos\alpha~sin\alpha
\label{eq10.2}
\medskip
\end{eqnarray}
 where $cos\alpha~=~\hat{p}\cdot\hat{q}$. It should be pointed
out that for a nucleus with an arbitrary $J$  the function
$B^{N}_{0,{\cal{M}}}$ (for $J~\geq~1$) and $B^{N}_{1(2),{\cal{M}}}$ ((for
$J~>~1$)) depend upon even powers of the pseudoscalar quantity ($\vec{S}_A\cdot
\hat{p}$) as well (see Appendix A).

\indent The evaluation of $G^1_A$ and $G^2_A$ within the PWIA can be
carried out by substituting in Eqs. (\ref{5.1}) and (\ref{5.2}) the elements of
the PWIA hadronic tensor $w^{a}_{\mu\nu}$ obtained from
Eqs. (\ref{eq6}), (\ref{7}) and (\ref{eq7.3}), and chosing
$\hat{q}\equiv\hat{S}_A \equiv \hat{u}_z$, since the polarized structure
functions do not depend upon the direction of both $\vec{q}$ and $\vec{S}_A$.
In
order to use  ${\hbox{\Large\it P}}_{\parallel(\perp)}^{N}\!\left(
|\vec{p}|,E,\alpha\right)$ instead of the functions $B^{N}_{1(2),{1  \over
2}}~$  one can invert Eqs. (\ref{eq10.1}) and (\ref{eq10.2}); then one gets
( in what follows $p~\equiv~|\vec{p}|$ )

\begin{eqnarray}
{}~~~~\frac{G^A_1\left (Q^2,\nu\right )}{M_A}=2\pi
 \sum\nolimits\limits_{N=p,n}
\int \limits_{E_{min}} \limits^{E_{max}(Q^2,\nu)}dE\int
\limits_{p_{min}(Q^2,\nu,E)}\limits^{p_{max}(Q^2,\nu,E)}{p\over{|\vec{q}|E_p}}dp~
\left \{ \tilde{G}_{1}^{N}\!\left({Q}^{2}
\right) \left[M~{\hbox{\Large\it
P}}_{\parallel}^{N}\!\left( p,E,\alpha\right)~+
\right. \right. \nonumber \\
\left.\left.\;-p~\left({\nu\over|\vec q|}~
-{p~cos\alpha\over{M+E_p}}\right)
{}~\hbox{$\Large\cal P$}^{N}\!\left({p}
, E,\alpha\right)\right]~-~{Q^2\over|\vec q|^2}~{\hbox{$\Large\cal
L$}^N} \right \}~~~~~~~
 \label{eq11}
\end{eqnarray}

\begin{eqnarray}
{}~~~~\frac{G^A_2\left (Q^2,\nu\right )}{M_A^2}=2\pi
  \sum\nolimits\limits_{N=p,n}
\int \limits_{E_{min}} \limits^{E_{max}(Q^2,\nu)}dE\int
\limits_{p_{min}(Q^2,\nu,E)}\limits^{p_{max}(Q^2,\nu,E)}{p\over{|\vec{q}|E_p}}dp~
\left\{ \left[\tilde{G}_{1}^{N}\!\left({Q}^{2} \right)~{p\over
|\vec{q}|}{\hbox{$\Large\cal
P$}^{N}\!\left({p},E,\alpha\right)}\;+\right.\right.
\nonumber \\
\left.\left.+~{\tilde{G}_{2}^{N}\!\left({Q}^{2}
\right)\over M} \left(E_p~{\hbox{\Large\it
P}}_{\parallel}^{N}\!\left({p},E,\alpha\right)~-~{p^2~cos\alpha\over{M+E_p}}
{}~\hbox{$\Large\cal P$}^{N}\!\left({p} ,
E,\alpha\right)\right)\right]~-~{\nu\over|\vec
q|^2}~{\hbox{$\Large\cal
L$}^N} \right \}~~~~~~
\label{eq12}
\end{eqnarray}

\noindent with

\begin{eqnarray}
{\hbox{$\Large\cal
L$}^N}&=&\left[\tilde{G}_1^{N}\!\left ( Q^{2} \right)~{\cal
H}_1^N~+~|\vec q|~{\tilde{G}_{2}^{N}\!\left ( Q^{2} \right ) \over M}~{\cal
H}_2^N\right ]
\label{12.0}
\\
{\cal H}_1^N&=&{1\over
2}~{(3~cos^2\alpha~-~1)\over
cos\alpha}~\left[{p^2\over{M+E_p}}{\hbox{$\Large\cal P$}^{N}\!\left(
{p},E,\alpha\right)}~+~M~{{\hbox{\Large\it
P}}_{\perp}^{N}\!\left({p},E,\alpha\right)\over sin\alpha}\right]~~~~~~
\label{12.1}
\\
{\cal H}_2^N&=&p~\left[{\hbox{$\Large\cal
P$}^{N}\!\left( {p},E,\alpha\right)}~-~{{\hbox{\Large\it
P}}_{\perp}^{N}\!\left({p},E,\alpha\right)\over
sin\alpha}\right]+ \nonumber \\
&~&-~{\nu\over
2|\vec{q}|}~{(3~cos^2\alpha~-~1)\over
cos\alpha}~\left[{p^2\over{M+E_p}}{\hbox{$\Large\cal P$}^{N}\!\left(
{p},E,\alpha\right) }~-~E_p~{{\hbox{\Large\it
P}}_{\perp}^{N}\!\left({p},E,\alpha\right)\over sin\alpha}\right] \label{12.2}
\medskip
 \end{eqnarray}
and ${\hbox{$\Large\cal P$}}^{N}\!\left( {p},E,\alpha\right) =
cos\alpha~{\hbox{\Large\it
P}}_{\parallel}^{N}\!\left({p},E,\alpha\right)~+~sin\alpha~{\hbox{\Large\it
P}}_{\perp}^{N}\!\left({p},E,\alpha\right)$. In Eqs. (\ref{eq11}) and
(\ref{eq12}) the integration limits and $cos \alpha$ are determined, as usual,
by energy conservation \cite{9.}. The polarized structure functions
$G^A_{1(2)}$, given by Eqs. (\ref{eq11}) and (\ref{eq12}), coincide with the
ones
corresponding to the extraction scheme (A) of Ref. \cite{6.}, once the
off-shell
effects are neglected and
 ${\hbox{\Large\it
P}}_{\parallel(\perp)}^{N}\!\left({p},E,\alpha\right)$ are expressed in terms
of the scalar functions $f_1$ and $f_2$ introduced in Ref. \cite{6.}.

\indent It should be pointed out that the expression for $G^A_{1(2)}$, obtained
in Ref. \cite{5.},
 differ from Eqs. (\ref{eq11}) and (\ref{eq12})
in that they do not contain the term ${\hbox{$\Large\cal
L$}^N}$. The origin of such a difference can be traced back to the procedure
of Ref. \cite{4.}, used in Ref. \cite{5.}, according to which the polarized
structure functions are obtained from  Eqs. (\ref{5.3}) and (\ref{5.4}) by
replacing the four-vector $V^\sigma$  with the four-vector
$R^\sigma$, i.e. according to prescription II. As already explained in Sect.2,
such a procedure is a correct one only if the functional dependence of the two
four-vectors is the same, i.e. if $R^\sigma$ is a linear combination of only
$S^\sigma_A$ and  $P^\sigma_A$. It turns out by an explicit evaluation of Eq.
(\ref{7}) that this is not the case, for $R^\sigma$  is a linear combination of
$S^\sigma_A$,   $P^\sigma_A$ and  a term proportional to  $q^\sigma$. A
different situation occurs when, instead of the vectors  $V^\sigma$ and
$R^\sigma$, the tensors $W^a_{\mu\nu}$ and $w^a_{\mu\nu}$ are considered and
Eqs.
(\ref{5.1}) and (\ref{5.2}) are used. As a matter of fact, as already observed,
Eqs. (\ref{5.1}) and (\ref{5.2}) hold independently of the presence of a term
proportional to $q^\sigma$, since it is washed out by
$\epsilon_{\mu\nu\alpha\sigma}$$q^\alpha$ when the
 hadronic tensor is evaluated. Then the tensor
$W^a_{\mu\nu}$ can be safely replaced by $w^a_{\mu\nu}$ in Eqs. (\ref{5.1}) and
(\ref{5.2}).

Let us  now
analize in more detail the differences between prescription I and II. By
recalling that from the knowledge of the hadronic tensor follows the knowledge
of the four-vector  $\tilde{V}^\sigma~\equiv~V^\sigma~+  {q^\sigma\over
Q^2}~(V\cdot q)$ only,
 we can identify such a four-vector with its PWIA counterpart, i.e.
$\tilde{R}^\sigma~\equiv~R^\sigma~+ {q^\sigma\over Q^2}~(R\cdot q)$. Then we
can  express Eqs. (\ref{5.1}) and (\ref{5.2}) in terms of $\tilde{R}^\sigma$
and
eventually in terms of $R^\sigma$. One gets
 \begin{eqnarray} {G_1^A\over M_A}&=&-{(R\cdot q)\over
|\vec{q}|S_{Az}}~-~{Q^2\over |\vec{q}|^2} \left({R_z\over S_{Az}}~-~
{R_\perp\over S_{A\perp}}\right)   \label{13.1} \\  {G_2^A\over
M_A^2}&=&{R_0\over|\vec{q}|S_{Az}}~-~{\nu\over |\vec{q}|^2} \left({R_z\over
S_{Az}}~-~ {R_\perp\over S_{A\perp}}\right)  \label{13.2}
\medskip
\end{eqnarray}
where  $R_{\perp}=(\vec{R} \cdot
\vec{S}_A-R_zS_{Az})/S_{A\perp}$. From these equations, after inserting the
actual expression for $R^\sigma$ ( Eq. (\ref{7}) ), one gets again
Eqs.(\ref{eq11}) and (\ref{eq12}). It should be pointed out that
Eqs. (\ref{13.1}) and (\ref{13.2}) reduce to Eqs.(\ref{5.3}) and (\ref{5.4})
if the term $\left({R_z/ S_{Az}}~-~ {R_\perp/ S_{A\perp}}\right)$ vanishes.
However this is not the case, because  $\vec R$ does not result to be parallel
to $\vec S_A$.

\bigskip

\noindent{\bf 4. The asymmetry in the quasi-elastic region}
 \bigskip

The contraction of
the two tensors in Eq. (\ref{eq1}) yields

\begin{eqnarray}
\frac{d^2\hbox{$\large\sigma$}(h)}{d\Omega_2 d\nu}~=~
\hbox{$\large\Sigma$}\;+\;h\;\hbox{$\large\Delta$}
\label{eq14}
\end{eqnarray}

where

\begin{eqnarray}
\hbox{$\large\Sigma$}~=~
\hbox{$\large\sigma$}_{Mott} \left
[W^A_2(Q^2,\nu)\;+\;2~tan^2\frac{\theta_{e}}{2}~W^A_1(Q^2,\nu)\right]
\label{eq14.1}
\end{eqnarray}
\begin{eqnarray}
\hbox{$\large\Delta$}~=~
\hbox{$\large\sigma$}_{Mott}~2~tan^2\frac{\theta_{e}}{2}
\left[\frac{G^A_1(Q^2,\nu)}{M_A}~(\vec{k_1}+\vec{k_2})~+~2~
\frac{G^A_2(Q^2,\nu)}{M_A^2}~(\epsilon_1
 \vec{k_2}-\epsilon_2 \vec{k_1})\right]\cdot\vec{S_A}
\label{eq14.2}
\medskip
\end{eqnarray}
with $\theta_{e}$ being the scattering angle.
In
what follows, the target polarization vector $\vec{S_A}$ is supposed to lie
within the scattering plane formed by $\vec k_1$ and $\vec k_2$.

\indent Two possible kinematical conditions can be considered:

\indent{\em{$\beta$ - kinematics}}. The target polarization angle is measured
with respect to the direction of the incident electron, i.e. $cos\beta =
\vec{S_A} \cdot \vec{k_1}/|\vec{k_1}|$, (this is a natural choice from
the experimental point of view). In this case, one gets

\begin{eqnarray}
\hbox{$\large\Delta$}~\equiv~\hbox{$\large\Delta$}_\beta&=&
\hbox{$\large\sigma$}_{Mott}~2~tan^2\frac{\theta_{e}}{2}
\left\{\frac{G^A_1(Q^2,\nu)}{M_A}~\left[\epsilon_1 cos\beta\;+\;\epsilon_2
cos(\theta_{e}-\beta)\right]~+\right. \nonumber \\
&~&\left. -~2~
\frac{G^A_2(Q^2,\nu)}{M_A^2}~\epsilon_1 ~\epsilon_2 \left[cos\beta -
cos(\theta_{e}-\beta)\right]
\right\}
\label{eq16}
\medskip
\end{eqnarray}

\indent{\em{$\theta^*$ - kinematics}}. The target polarization angle is
measured with respect to the direction of the momentum transfer, i.e.
$cos\theta^* = \vec{S_A} \cdot \vec{q}/|\vec{q}|$. Then one can write

\begin{eqnarray}
\hbox{$\large\Delta$}~\equiv~\hbox{$\large\Delta$}_{\theta^*}&=&
-\hbox{$\large\sigma$}_{Mott}~tan\frac{\theta_{e}}{2}
\left\{cos\theta^*~R^A_{T'}(Q^2,\nu)~\left[{Q^2 \over |\vec{q}|^2} +
tan^2\frac{\theta_{e}}{2}\right]^{1/2}~+ \right.
\nonumber \\
&~&\left. -~\frac{Q^2}
{|\vec{q}|^2~\sqrt{2}}~sin\theta^* ~R^A_{TL'}(Q^2,\nu)\right\}
\label{eq17}
\medskip
\end{eqnarray}
where
\begin{eqnarray}
R^A_{T'}(Q^2,\nu)&=& -2~ \left(\frac{G^A_1(Q^2,\nu)}{M_A}~ \nu~ - ~Q^2
\frac{G^A_2(Q^2,\nu)}{M_A^2}\right)
{}~=~i~2~{W^{a}_{12}\over S_{Az}}
 \label{eq17a}
\end{eqnarray}
\begin{eqnarray}
R^A_{TL'}(Q^2,\nu)&=&2~\sqrt{2}~|\vec{q}|~\left(\frac{G^A_1(Q^2,\nu)}{M_A}~
+~\nu
\frac{G^A_2(Q^2,\nu)}{M_A^2}\right)
{}~=~-i~2\sqrt{2}~{W^{a}_{02}\over S_{Ax}}
\label{eq17b}
\end{eqnarray}
In principle the  $\theta^*$ - kinematics is very appealing, since by
performing experiments at $\theta^* = 0$ and $90^o$ one can disentangle
$R^A_{T'}$ and $R^A_{TL'}$, which, at the top of the qe peak, are proportional
to  $(G^n_M)^2$ and
$G^n_E~G^n_M$, respectively, provided the proton contribution can be
disregarded \cite{2.,3.}. Let us analize in what follows to what extent such
a condition really occurs.

\indent
Experimentally one measures the longitudinal asymmetry defined as

\begin{eqnarray}
\rm A= \frac {{\sigma }_{2}\left({\nu ,{Q}^{2},{\vec{S}}_{A},+1}\right) -
{\sigma
}_{2}\left({\nu ,{Q}^{2},{\vec{S}}_{A},-1}\right)} {{\sigma }_{2}\left({\nu
,{Q}^{2},{\vec{S}}_{A},+1}\right) + {\sigma }_{2}\left({\nu
,{Q}^{2},{\vec{S}}_{A},-1}\right)} =
\frac{\hbox{$\large\Delta$}}{\hbox{$\large\Sigma$}}
 \label{eq18}
\end{eqnarray}
If the naive model of $^3$He holds, this quantity is in principle very
sensitive
to the neutron properties, since the numerator should be essentially given by
the
neutron with its spin aligned along $\vec S_A$. With this simple picture
in mind, let us consider the comparison between our  results based on
Eqs.(\ref{eq11}) and (\ref{eq12}), with the experimental data obtained at
MIT-Bates \cite{2.,3.}.

\indent In Fig. 3 the asymmetry corresponding to $\epsilon_1 = 574~MeV$ and
$\theta_e=44^o$,   measured  by the
MIT-Caltech collaboration \cite{2.} is shown. The experimental data were
obtained in a large interval of energy transfer after averaging over three
different values of the $\beta$ angle ($\beta=44.5^o, 51.5^o,135.5^o$, with the
corresponding azimuthal angles being: $\phi=180^o,180^o,0^o$). It is worth
noting
that in these kinematical conditions one has $\theta^*~\approx~90^o$
only at the top of the qe peak, and  therefore only there the measured
asymmetry reduces to $R_{TL'}$. In the figure  the neutron (dotted line) and
proton (dashed line) contributions are separately  shown, and  the
relevance of the proton  contribution can be noticed  particularly at the top
of
the qe peak. There, one has

\begin{eqnarray}
A^{th}& = & 3.74~\% \nonumber\\
A^{th}_p & = & 2.20~\%
\nonumber
\medskip\end{eqnarray}
for the Gari-Kruempelmann \cite{12.} form factors of the nucleon. Similar
results hold for other models of the nucleon form factors, such
as the Blatnick-Zovko \cite{14.} and the Hoehler \cite{15.} ones. In order to
compare the theoretical prediction with the experimental data at the qe peak,
one has to perform a further averaging over an interval of the energy transfer
of about $100~MeV$, as it was done in the experiment of Ref. \cite{2.}; then
one
gets \begin{eqnarray}
A^{exp}_{qe}& = & 2.41~\mp1.29~\mp0.51~\%  ~~ MIT-Caltech ~{\cite{2.}}
\nonumber\\
A^{th}& = & 1.65~\% \nonumber\\
A^{th}_p & \simeq & 0
\nonumber
\medskip\end{eqnarray}
The aim of this kind of experiments is the extraction of information on the
 neutron form factors $G_{E}^n$
and $G_M^n$ from the experimental asymmetry. The last comparison and the
theoretical curves shown in Fig. 3 suggest that, since the asymmetry
drastically
changes when the energy transfer varies, the averaging procedure has to be
considered with some care. Indeed, the above mentioned proportionality between
$R_{TL'}$ and
$G_{E}^n~G_M^n$ holds only at the top of the qe peak and provided the proton
contribution is negligible, whereas the theoretical calculations show that
the proton contribution is relevant at the
top of the qe peak. This
result is completely hidden by the averaging procedure and therefore the
extraction of information on the neutron form factors from
the  experimental data averaged on a large energy range could be questionable.

 For the sake of completeness, herebelow we compare our results for the
asymmetry at the qe peak with the experimental results of Ref.\cite{3.}
($\epsilon_1 =
578~MeV$, $\theta_e=51.1^o$ and $\theta^*=90.2$):
\begin{eqnarray} A^{exp}_{qe} &
= & 1.75~\mp1.20~\mp0.31~\%  ~~ MIT-Harvard~{\cite{3.}} \nonumber\\
A^{th}&=&4.00~\% \nonumber\\ A^{th}_p&=&2.78~\%  \nonumber
\medskip\end{eqnarray}
It should be pointed out that for this experiment the
energy transfer range of the experimental averaging has not been specified.

\indent In Fig.4 the theoretical asymmetry, corresponding to
$\epsilon_1~=574~MeV$ and  $\theta_e~=~51.1^o$ and averaged over the same
values
of the polarization angle as in Fig. 3, is shown. Such a kinematics
was chosen \cite{2.} with the aim of extracting $R_{T'}$ at the qe peak. Only
one  experimental point has been obtained  for the averaged asymmetry around
the
top of the qe peak, where $\theta^*~\approx~0^o$
($A^{exp}_{qe}\propto~R_{T'}^{exp}$). The comparison between the experimental
result (obtained after a further averaging over an interval of
the energy transfer of  $48~MeV$) and our calculations is as follows

\begin{eqnarray}
A^{exp}_{qe}&=&-3.79~\mp1.37~\mp0.67~\%~~MIT-Caltech ~ {\cite{2.}}
\nonumber\\
A^{th}&=&- 4.30~\% ~~~~~~[ without~averaging~~-3.43~\%] \nonumber\\
A^{th}_p&=&~1.05~\%~~~~~~ [ without~ averaging~~ 1.30~\%]
\nonumber
\medskip\end{eqnarray}
Our calculation without averaging compare to the response function $R_{T'}$
corresponding to the kinematics of the experiment of Ref. \cite{3.}
($\epsilon_1 =
578~MeV$ and $\theta_e=51.1^o$, and two values of the
polarization angles: $\theta^*=3.2^o$, $\phi^*=0$ and
$\theta^*=176.8^o$, $\phi^*=180^o$)  as follows
\begin{eqnarray} \nonumber\\
A^{exp}_{qe}&=&-2.60~\mp0.90~\mp0.46~\%~~MIT-Harvard~{\cite{3.}}
\nonumber\\
A^{th}&=&-3.68~\% \nonumber\\
A^{th}_p&=&1.18~\%
\nonumber
\medskip\end{eqnarray}
 The same comments and warnings made for the case $\theta^*=90^0$ should be
extended to $\theta^*=0^o$ as well. It should be pointed out that our numerical
results, as shown in Fig. 5, only slightly differ from the ones obtained
in Ref. \cite{6.}, where a spin-dependent Faddeev spectral function has been
used.

\indent In Figs. 6a and 6b  the results based upon prescription
I are compared with  our  previous calculations \cite{5.}, based upon
prescription II (the corresponding explicit expressions for $G^A_{1(2)}$ are
given in Ref. \cite{5.} and coincide, as already mentioned, with Eqs.
(\ref{eq11}) and (\ref{eq12}) with the term ${\hbox{$\Large\cal L$}^N}$
dropped
out). The results of the comparison show that at
$\theta^*~\approx~90^o$ prescription II yields results very different from
the correct ones, whereas at $\theta^*~\approx~0^o$
 such a difference is not present.   This is due to the fact that in
procedure II, based
on  Eqs. (\ref{5.3}) and (\ref{5.4}), $R_{TL'}$,
Eq. (\ref{eq17b}), results to be proportional to the component of $\vec{R}$
along
$\hat{q}$, instead of being proportional to its transverse part (i.e.
$W^A_{02}\propto {R_\perp / S_\perp}$, see Eqs. (\ref{13.1}),  (\ref{13.2}) and
(\ref{eq17b})), as it should be. Therefore $R_{TL'}$, within prescription II,
is affected by the presence of a term
 proportional to $q^\sigma$ in
 $R^\sigma$.  For
$\theta^*~\approx~0^o$ the differences, as shown in Fig. 6b, are very small
over
the whole range of the energy transfer considered, since $R_{T'}$ is the same
for the two prescriptions, and therefore it is unaffected by the extra term
in$~R^\sigma$. Moreover it is easy to explain the small
 differences on the wings of the
asymmetry, since the asymmetry is proportional to  $R_{T'}$ only at the
top of the qe peak, while on the wings there is a  mixing with $R_{TL'}$.

\noindent From the above comparisons it turns out that the difference between
the two procedures is almost entirely due to the proton contribution;
nevertheless, the correctness of our conclusion, reached in \cite{5.} using
prescription II, about the possibility of obtaining information on the neutron
form factors by properly minimizing the proton contribution is not affected by
the use of  prescription I. This will be illustrated in the next Section.

\bigskip
\noindent{\bf 5. Minimizing the proton contribution}
\bigskip

\indent As shown in Figs. 3 and 4, the proton contribution to the asymmetry,
corresponding to the polarization angle of the actual experiments is sizeable.
Following our previous paper \cite{5.}, in this Section the possibility to
minimize or even to make vanishing the proton contribution will be
investigated,
using prescription I. To this end we have analyzed the proton contribution to
the
asymmetry at the top of the qe peak, for different values of $\beta$, different
values of the energy of the incident electron and different models for the
proton form factors. The results are presented in Fig. 7. It can be seen that
for $\theta_e~=~75^o$
the proton contribution is almost vanishing around $\beta\equiv\beta_c=105^o$,
in a large  spectrum of values of incident electron energy; moreover, such a
feature  weakly depends upon the model for the nucleon form factors. In order
to
understand the behaviour of the proton contribution, let us consider the
asymmetry given by Eq. (\ref{eq18}). The proton contribution vanishes in
(\ref{eq18}) if in $
\hbox{$\large\Delta$}_\beta$, Eq. (\ref{eq16}), a polarization angle
$\beta=\beta_c$  is chosen such that

\begin{eqnarray}
\tan\beta_c~=~-{1\over \tan\theta_e}~-~{1\over
\sin\theta_e}~\gamma
\label{eq20}
\medskip
\end{eqnarray}
with

\begin{eqnarray}
\gamma~=~{\left ( G^{A(p)}_{1} M_A/(2\epsilon_2~
G^{A(p)}_{2})~-~ 1 \right ) \over {\left (G^{A(p)}_{1} M_A/(2\epsilon_1~
G^{A(p)}_{2})~+~1\right )}}
\label{eq21}
\medskip
\end{eqnarray}
The critical polarization angle does not change too much, even if $\gamma$
changes by orders of magnitude, since the equation determining $\beta_c$
involves
the {\em tangent} function, which always results to be $\mid \tan \beta_c \mid
\gg 1$. As a matter of fact, we have checked that for
$10^o~\leq~\theta_e~\leq~80^o$ and $.5~(GeV)~\leq~\epsilon_1~\leq~3~(GeV)$,
$\beta_c$ varies between $89^o$ and $110^o$. This is due to the fact that
$\gamma$ is always large ($\mid\gamma\mid >> 1)$, since $G^{A(p)}_{1} M_A/
G^{A(p)}_{2}$ is negative and the denominator in Eq. (\ref{eq21}) is small for
all nucleon form factor models we have used.
Therefore the presence of the {\em tangent} function in Eq. (\ref{eq20})
explains the weak dependence of $\beta_c$ upon sizeable changes of  both the
incident energy, the scattering angle and the models of the nucleon form
factors.
 In Fig. 8,  the
asymmetry and the proton contribution, vs  $Q^2$, at fixed values of
$\beta_c~=~95^o$ and $\theta_e~=~75^o$ are presented for three different models
of the nucleon form factors (Refs. \cite{12.,14.,15.}). It can be seen that the
asymmetry is sensitive to the neutron form factors; moreover the calculations
have shown that the neutron asymmetry is essentially given by the terms in
$G^{A}_{1(2)} $, containing ${\hbox{\Large\it
P}}_{\parallel}^{n}\!\left( |\vec{p}|,E,\alpha\right)$.
However the results presented in Fig. 8 do not tell us whether the
differences in the asymmetry are given by the differences in   $G_E^n$ or in
$G^n_M$, since both of them vary within the  models we have considered. In
order
to make our analysis a more stringent one, we have repeated the calculation by
using the Galster model of the nucleon form factors \cite{13.}, since  within
such a model  $G_E^n$ can be changed independently of $G_M^n$. In fact one has

\begin{eqnarray}
G_M^n&=&{\mu_n~G_E^p} \nonumber \\
G_E^n&=&{-\tau~\mu_n\over(1+\eta~\tau)}~G_E^p
\label{19}
\medskip
\end{eqnarray}
where  $G_E^p = 1/(1+Q^2/B)^2$, $B~=~0.71 (GeV/c)^2$ and $\eta$ is a parameter.
The resulting asymmetry and proton contribution are shown in Fig. 9 for
different values of  $\eta$.  Fig. 9
illustrates  how the total asymmetry can depend upon  $G^n_E$, having, at the
same time, a vanishing proton contribution.

\noindent It should moreover be stressed that the
proposed kinematics, which minimizes the proton contribution, corresponds to
the  qe peak, where the final state interaction is expected to play a minor
role.

\bigskip
\noindent{\bf 6. Summary and conclusion}
\bigskip

\indent The qe spin-dependent structure functions for a nucleus with J =1/2
have
been obtained by a proper procedure, based on the replacement of the exact
hadronic tensor with its PWIA version. Our formal results are in agreement with
the ones of Ref. \cite{6.}, and the numerical calculations only slightly differ
from
them, which demonstrates that the spin dependent spectral
functions used in Ref. \cite{5.} and  the one used Ref. \cite{6.} are
essentially equivalent.

\indent The differences between the predictions of the correct
procedure and the ones \cite{4.,5.} based upon the replacement of the hadronic
pseudovector $V^\sigma$, Eq.(\ref{eq3.3}), with its PWIA version $R^\sigma$,
Eq.(\ref{7}), have  been shown to be produced by the presence of a
contribution proportional to the momentum transfer $q^\sigma$ in the
four-vector
$R^\sigma$. This contribution affects only the response function $R_{TL'}$.

\indent Our analysis of the asymmetry, based on the correct expression
of $G^A_1$ and $G^A_2$ given by Eqs. (\ref{eq11}) and
(\ref{eq12}), respectively, has fully confirmed the main conclusions of our
previous paper \cite{5.}, concerning : i) the relevance of the proton
contribution for the experimental kinematics  considered till now, and ii) the
possibility to select a polarization angle, which leads at the qe peak to an
almost vanishing proton contribution for a wide range of kinematical variables;
within such a kinematical condition, the sensitivity of the asymmetry to the
electric neutron form factor has been thoroughly investigated.

\indent Calculations of the final state effects are in progress.

\bigskip
\noindent{\bf 7. Acknowledgment}
\bigskip

\noindent We would like to acknowledge valuable discussions with F. Coester, C.
E. Jones,
 R.D. McKeown, R. Milner, P.U. Sauer and R.W. Schulze. We are
indebted to C.E. Jones  for providing us with detailed information on the
experiment of Ref. \cite{2.}.

\bigskip
\noindent{\bf Appendix A}

\bigskip

\indent For the sake of completeness, let us repeat here Eqs.(\ref{eq9.1}) and
(\ref{eq10})
 for a nucleus with an arbitrary value of the total angular momentum $J_A$
 \begin{eqnarray}
{\bf\hat{P}}^{N}_{\cal{M}}(\vec{p},E)={1\over
2}\left[B_{0,{\cal{M}}}^{N}(|\vec{p}|,E,(\widehat{ p} \cdot
\widehat{S}_A)^2)~+~\vec{\sigma} \cdot \vec{f}^{N}_{\cal
{M}}(\vec{p},E)\right]
\label{A0A}
\medskip
\end{eqnarray}

\begin{eqnarray}
\vec{f}^{N}_{\cal{M}}(\vec{p},E)~=~\vec{S}_A~B_{1,\cal{M}}^{N}(|\vec{p}|,E,(\widehat{ p} \cdot
\widehat{S}_A)^2)~+~\hat{p}~(\hat{p}
\cdot \vec{S}_A)~ B_{2,\cal{M}}^{N}(|\vec{p}|,E,(\widehat{ p} \cdot
\widehat{S}_A)^2)
\label{A0B}
\medskip
\end{eqnarray}

\noindent In
this Appendix the expressions  of the functions  $B^{N}_{0(1,2),\cal{M}}$, will
be given in terms of the overlap
integrals \cite{8.}

 \begin{eqnarray}
_{N}\langle
\vec{\rho},\sigma~;\psi^{J_{A-1},M_{A-1}}_
{E_{f_{{(A-1)}}},\eta}|\Psi^{J_{A},M_{A}} \rangle \label{A0C}
\medskip
\end{eqnarray}
where $\psi^{J_{A-1},M_{A-1}}_{E_{f_{{(A-1)}}},\eta}$ is the wave function of
the (A-1) spectator system, $ \Psi^{J_{A},M_{A}}$ the wave function of the
target nucleus, $\sigma$,
$M_{A-1}$ and $M_{A}$ are the components of the angular momentum of the nucleon
$N$, the (A-1) system and the target system along $\hat{q}$, respectively, and
$\vec{\rho}$ is the Jacobi coordinate of the nucleon $N$ with respect to the
(A-1) system, while $\eta$ represents all of the quantum numbers of the (A-1)
system, but the energy  $E_{f_{(A-1)}}=E~+~E_A$ and the total angular momentum
$J_{A-1}$ with its third component $M_{A-1}$.

The
starting point is the usual transformation property from a wave function with a
given third component of the total angular momentum with respect to a certain
axis (e.g. with respect to the $\hat{S}_A$-axis) to the wave function
where the third component of the angular momentum is defined with respect to
another axis  (e.g. the $\hat{q}$-axis). By using the Wigner D-function
$D^{J_A}_{M_A{\cal{M}}}$ one has \cite{16.}
 \begin{eqnarray}
\langle\theta\phi|J_A~{\cal{M}}\rangle_{\hat{S}_A}~=~
\langle\theta\phi|D|J_A~{\cal{M}}\rangle_{\hat{q}}~=~
\sum_{M_A}~\langle\theta\phi|J_A~M_A\rangle_{\hat{q}}~D^{J_A}_{M_A{\cal{M}}}(\alpha,\beta,\gamma)
\label{A00} \medskip \end{eqnarray}
where  the subscripts $\hat{q}$ and $\hat{S}_A$ indicate the angular momentum
quantization axis, and $\alpha$, $\beta$, $\gamma$ are the Euler angles
describing the proper rotation from $\hat{q}$ to $\hat{S}_A$. This
transformation has to be applied to the overlap integrals
$_{N}\langle{\vec{p},\sigma;\psi }_{f_{(A-1)}} |{\psi }_{J\cal{M}}\rangle$,
which appear in the definition of the spin dependent spectral function (see Eq.
(\ref{eq9}) ). Then, after a lenghty algebraic manipulations and using the
following properties of the bipolar spherical harmonics, which hold for odd
values of $\jmath$, \begin{eqnarray}  \sum_{\mu\lambda}~\langle 1a, (\jmath+1)
\lambda|\jmath \mu \rangle~{\cal{Y}}_{\jmath}^{\mu~\ast}\left(\widehat{S}_A
\right)~{\cal{Y}}_{\jmath+1}^{\lambda}\left(\widehat{p}\right)~=~-{\widehat{\jmath}
\over \sqrt{3~(\jmath +1)}}~~~\bullet~~~~~~~~~~~~~~ \nonumber \\
\bullet~\left[{\cal{Y}}_{1}^{a\ast}\left(\widehat{p}
\right)~\sum_{\ell=0}^{\jmath-1 \over 2}~{\widehat{( 2 \ell+1)}}~{\cal{Y}}_{2
\ell+1}^0\left(\widehat{p}\cdot \widehat{S}_A
\right)~-~{\cal{Y}}_{1}^{a\ast}\left( \widehat{S}_A
\right)~\sum_{\ell=0}^{\jmath-1 \over 2}~{\widehat{( 2 \ell)}}~{\cal{Y}}_{2
\ell}^0\left(\widehat{p}\cdot \widehat{S}_A \right)\right] \nonumber \\
\nonumber \\ \sum_{\mu\lambda}~\langle 1a, (\jmath-1) \lambda|\jmath \mu
\rangle~{\cal{Y}}_{\jmath}^{\mu~\ast}\left(\widehat{S}_A
\right)~{\cal{Y}}_{\jmath-1}^{\lambda}\left(\widehat{p}\right)~=~-{\widehat{\jmath}
\over \sqrt{3~\jmath }}~~~\bullet~~~~~~~~~~~~~~ \nonumber \\
\bullet~\left[{\cal{Y}}_{1}^{a\ast}\left(\widehat{p}
\right)~\sum_{\ell=0}^{\jmath-3 \over 2}~{\widehat{( 2
\ell+1)}}~{\cal{Y}}_{2 \ell+1}^0\left(\widehat{p}\cdot \widehat{S}_A
\right)~-~{\cal{Y}}_{1}^{a\ast}\left( \widehat{S}_A
\right)~\sum_{\ell=0}^{\jmath-1 \over 2}~{\widehat{( 2
\ell)}}~{\cal{Y}}_{2 \ell}^0\left(\widehat{p}\cdot \widehat{S}_A
\right)\right]
\label{a01}
\medskip
\end{eqnarray}

one gets
   \begin{eqnarray}
B^{N}_{0,{\cal{M}}}\left(|\vec{p}|,E,(\widehat{ p} \cdot
\widehat{S}_A)^2\right)~=~ {\widehat{J}_A \sqrt{2} \over
\pi^{3/2}}~\sum_{\jmath_{even} =0}^{2J_A}~\widehat{\jmath}~\langle J_A
{\cal{M}}
\jmath 0|J_A {\cal{M}}\rangle~{\cal{Y}}_{\jmath}^0\left(\widehat{p}\cdot
\widehat{S}_A \right)~F^{J_A}_{N}(\jmath,\jmath,0)  \label{A0}  \end{eqnarray}
  \begin{eqnarray}
 B^{N}_{1,{\cal{M}}}\left(|\vec{p}|,E,(\widehat p \cdot
\widehat{S}_A)^2\right)~=~ {\widehat{J}_A \over
\pi^{3/2}}~\sqrt{6}~\sum_{\jmath_{odd} =1}^{2J_A}~
\widehat{\jmath}~\langle
J_A {\cal{M}} \jmath 0|J_A
{\cal{M}}\rangle~\bullet \nonumber \\
\bullet~\left [ {F^{J_A}_{N}(\jmath,\jmath +1,1) \over  \sqrt{\jmath+1 } }
 ~+~{F^{J_A}_{N}(\jmath,\jmath -1,1) \over \sqrt{\jmath } }
\right ]~\left [\sum_{\ell=0}^{\jmath-1 \over 2}~{\widehat{( 2
\ell)}}~{\cal{Y}}_{2 \ell}^0\left(\widehat{p}\cdot \widehat{S}_A
\right)  \right ] \label{A1} \medskip \end{eqnarray}
\begin{eqnarray}
B^{N}_{2,{\cal{M}}}\left(|\vec{p}|,E,(\widehat p \cdot
\widehat{S}_A)^2\right)~=~- {\widehat{J}_A \over (\widehat p \cdot
\widehat{S}_A)\pi^{3/2}}~\sqrt{6}~\sum_{\jmath_{odd}
=1}^{2J_A}~\widehat{\jmath}~\langle
J_A {\cal{M}} \jmath 0|J_A
{\cal{M}}\rangle~\bullet \nonumber \\ \bullet~\Biggl \{
\left [ {F^{J_A}_{N}(\jmath,\jmath+1,1) \over
\sqrt{\jmath+1}}  ~+~{F^{J_A}_{N}(\jmath,\jmath-1,1) \over \sqrt{\jmath
}}\right
]~\left [\sum_{\ell=0}^{\jmath -1 \over
2}~{\widehat{(2\ell+1)}}~{\cal{Y}}_{2\ell+1}^0\left(\widehat{p}\cdot
\widehat{S}_A \right)  \right ] ~+ \nonumber \\
-~{\cal{Y}}_{\jmath}^0\left(\widehat{p}\cdot
\widehat{S}_A \right)~F^{J_A}_{N}(\jmath,\jmath -1,1) \Biggr \}
{}~~~~~~~~~~~~~~~~\label{A2} \end{eqnarray}
with
\begin {eqnarray} F^{J_A}_{N}(\jmath,{\cal
{L}},a)~=~(-1)^{a}~\sum_{J_{A-1},\eta}~
\sum_{L,\tilde{L}}~\sum_{X,\tilde{X}}~(-1)^{1/2+X+J_{A-1}}~(-1)^{(L+\tilde{L})/2}~
\widehat{X} \widehat{\tilde{X}} ~\widehat{L}
\widehat{\tilde{L}}~~\bullet~~~~\nonumber \\  \bullet~
\langle L0\tilde{L}0|{\cal{L}}0 \rangle
\left \{ \begin{array}{ccc} {1/ 2}& 1/2 &a\\X&\tilde{X}&J_{A-1}\end{array}
\right
\}~\left \{ \begin{array}{ccc}  L&X&J_A\\\tilde{L}&\tilde{X}&J_{A}\\
{\cal{L}}&a& \jmath \end{array} \right \}
% \nonumber \\  \bullet
{}~{\hbox{\Large\it
I}}_{L~J_A~N}^{X~J_{A-1}~\eta}\!\left( |\vec{p}|, E\right)~{\hbox{\Large\it
I}}_{\tilde{L}~J_A~N}^{\tilde{X}~J_{A-1}~\eta}\!\left( |\vec{p}|, E\right)
%~~~~~~~~~~~~~~~~
\label{A3} \medskip
\end{eqnarray}
where $\widehat y$ means $\sqrt{2y+1}$ and $\jmath=0,
...,~2J_A$;  ${\cal L}= \jmath-1,\jmath,\jmath+1$; $a=0,1$. The quantity
${\hbox{\Large\it I}}_{L~J_A~N}^{X~J_{A-1}~\eta}\!\left( |\vec{p}|, E\right )$
is
given by
 \begin {eqnarray}  {\hbox{\Large\it
I}}_{L~J_A~N~}^{X~J_{A-1}~\eta}\!\left( |\vec{p}|, E\right)~=~\int~
d\vec{\rho}~ j_{L}(|\vec{p}|\rho)~\sum_{M_L,M_X}
\langle L M_L X M_X|J_A M_A \rangle~{\cal{Y}}^{M_L\ast}_L(\Omega_{\widehat
\rho})~\bullet \nonumber \\ \bullet~
\sum_{M_{A-1}\sigma}~\langle
J_{A-1} M_{A-1} {1 \over 2} \sigma |X M_X \rangle~
_{N}\langle \vec{\rho},\sigma
;\psi^{J_{A-1},M_{A-1}}_{E_{f_{(A-1)}},\eta}
|\Psi^{J_{A},M_{A}} \rangle.
\label{A4} \medskip \end{eqnarray}

\newpage
\begin{center}
FIGURE CAPTIONS
\end{center}

\noindent Fig. 1a. The proton unpolarized spectral function of
$^3$He versus the removal energy E and the nucleon momentum
$p~\equiv~|\vec{p}|$ (see text and appendix A).
\bigskip

\noindent Fig. 1b. The function $\mid B_{1,{1 \over 2}}^p \mid$  (see Eq.
(\ref{eq10}) and appendix A) versus the removal energy E and the nucleon
momentum
$p~\equiv~|\vec{p}|$.
\bigskip

\noindent Fig. 1c.  The function $\mid B_{2,{1 \over 2}}^p \mid$  (see Eq.
(\ref{eq10}) and
appendix A) versus the removal energy E and the nucleon momentum
$p~\equiv~|\vec{p}|$.
\bigskip

\noindent Fig. 2a. The same as  Fig. 1a, but for the neutron.
\bigskip

\noindent Fig. 2b. The same as Fig. 1b, but for the neutron.
\bigskip

\noindent Fig. 2c.  The same as Fig. 1c, but for the neutron.
\bigskip

\noindent Fig. 3.  The asymmetry in
$\theta^*$ kinematics corresponding to  $\epsilon_1$ = 574~$MeV$ and $\theta_e~
= ~44^o$,  vs. the energy transfer $\nu$. The experimental data are from
Ref. \cite{2.} and the theoretical curves were obtained using Eqs. (\ref{eq11})
and (\ref{eq12}) with the spin-dependent spectral function of Ref. \cite{5.}.
The continuous line is the total asymmetry, whereas the dotted (dashed) line
represents the neutron (proton) contribution. The nucleon form factors of Ref.
\cite{12.} have been used. The arrow indicates the position of the qe peak.
\bigskip

\noindent Fig. 4.
The same as in Fig. 3, but for $\theta_e~ =~ 51.1^o$. The experimental point
has
been obtained in Ref. \cite{2.}, after averaging over a 48 MeV interval around
the qe
peak.
\bigskip

\noindent Fig. 5. Comparison of the asymmetry, corresponding to $\epsilon_1$ =
574~$MeV$ and $\theta_e~ = ~44^o$ (cfr.
Fig. 3), calculated    by using Eqs. (\ref{eq11})
and (\ref{eq12}) and the spin-dependent
spectral function of Ref. \cite{5.} (solid line) with the one of Ref. \cite{6.}
(dashed
line), based on a Faddeev spin-dependent spectral function. The  nucleon form
factors of Ref. \cite{13.} have been used and the experimental data are from
Ref. \cite{2.}. The arrow indicates the position of the qe peak.

\bigskip

\noindent Fig. 6a. The asymmetry and the neutron contribution  for $\epsilon_1$
=
574 $MeV$ and $\theta_e~ =~ 44^o$   vs. the
energy transfer $\nu$ calculated using prescriptions I and II. Solid (dotted)
line: the
asymmetry (neutron contribution) corresponding to prescription I ( Eqs.
(\ref{eq11})
and (\ref{eq12}));  dashed (dot-dashed)
line: the asymmetry (neutron contribution) corresponding to
 prescription II ( Eqs. (\ref{eq11})
and (\ref{eq12}) without the term $\cal L$). The form factors of Ref.
\cite{12.} have been used and the experimental data are from
Ref. \cite{2.}. The  arrow indicates the position of qe peak.
\bigskip

\noindent Fig. 6b. The same as in Fig. 4a, but for  $\theta_e~ =~
51.1^o$. The dotted and dot-dashed lines overlap.
\bigskip

\noindent Fig. 7a. The proton  contribution
to the asymmetry , at the top of the qe peak , vs. $\beta$, for
$\theta_e~=~75^o$. Solid line: $\epsilon_1~=~500~MeV$, long-dashed line:
$\epsilon_1~=~1000~MeV$, short-dashed line: $\epsilon_1~=~1500~MeV$, dotted
line:
$\epsilon_1~=~2000~MeV$. The nucleon form factors of Ref. \cite{12.} have been
used. \bigskip

\noindent Fig. 7b. The same as in Fig. 7a, but for the nucleon form factors of
Ref. \cite{14.}. \bigskip

\noindent Fig. 7c. The same as in Fig. 7a, but for the nucleon form factors of
Ref. \cite{15.}.
\bigskip

\noindent Fig. 8. The total asymmetry  at the top of the qe
peak, vs. Q$^2$, for  $\theta_e~=~75^0$ and $\beta~=~95^o$, using Eqs.
(\ref{eq11})
and (\ref{eq12}). Solid line : Gari-Kruempelmann form factors \cite{12.};
dashed line:
Blatnik-Zovko form factors \cite{14.}; dotted line: Hoehler et al. form
factors \cite{15.}. The curves in the lower part of the figure represent the
corresponding proton contributions.
\bigskip

\noindent Fig. 9.  Sensitivity of the total asymmetry
upon variations of the neutron electric form factor. The curves in the upper
part were calculated at  $\theta_e~=~75^0$ and $\beta~=~95^o$, by Eqs.
(\ref{eq11})
and (\ref{eq12}) using the Galster form factors \cite{13.} for different values
of the
parameter  $\eta$. By this way the neutron electric form factor can be changed
leaving unchanged the magnetic form factor. The curves in the lower part of the
figure represent the corresponding proton contributions.

\newpage
\begin{figure}
%\special{picture b0p}
\vspace {14cm}
\end{figure}

$~~~$ \vspace{3cm}

Fig. 1a

\bigskip
C. CIOFI DEGLI ATTI, E. PACE and G. SALME'
\newpage
\begin{figure}
%\special{picture b1p}
\vspace {14cm}
\end{figure}

$~~~$ \vspace{3cm}

Fig. 1b

\bigskip
C. CIOFI DEGLI ATTI, E. PACE and G. SALME'
\newpage
\begin{figure}
%\special{picture b2p}
\vspace {14cm}
\end{figure}

$~~~$ \vspace{3cm}

Fig. 1c

\bigskip

C. CIOFI DEGLI ATTI, E. PACE and G. SALME'
\newpage
\begin{figure}
%\special{picture b0n}
\vspace {14cm}
\end{figure}

$~~~$ \vspace{3cm}

Fig. 2a

\bigskip
C. CIOFI DEGLI ATTI, E. PACE and G. SALME'
\newpage
\begin{figure}
%\special{picture b1n}
\vspace {14cm}
\end{figure}

$~~~$ \vspace{3cm}

Fig. 2b

\bigskip
C. CIOFI DEGLI ATTI, E. PACE and G. SALME'
\newpage
\begin{figure}
%\special{picture b2n}
\vspace {14cm}
\end{figure}

$~~~$ \vspace{3cm}

Fig. 2c

\bigskip
C. CIOFI DEGLI ATTI, E. PACE and G. SALME'
\newpage
\begin{figure}
\vspace{13cm}
%\special{picture prc3}
\vspace{2cm}
\end{figure}

$~~~$ \vspace{3cm}

 Fig.3

\bigskip
C. CIOFI DEGLI ATTI, E. PACE and G. SALME'

\newpage
\begin{figure}
\vspace{13cm}
%\special{picture prc4}
\vspace {2cm}
\end{figure}

$~~~$ \vspace{3cm}

Fig. 4

\bigskip
C. CIOFI DEGLI ATTI, E. PACE and G. SALME'
\newpage
\begin{figure}
\vspace{13cm}
%\special{picture prc5}
\vspace {2cm}
\end{figure}

$~~~$ \vspace{3cm}

Fig. 5

\bigskip
C. CIOFI DEGLI ATTI, E. PACE and G. SALME'
\newpage
\begin{figure}
\vspace{13cm}
%\special{picture prc6a}
\vspace {2cm}
\end{figure}

$~~~$ \vspace{3cm}

Fig. 6a

\bigskip
C. CIOFI DEGLI ATTI, E. PACE and G. SALME'
\newpage
\begin{figure}
\vspace{13cm}
%\special{picture prc6b}
\vspace {2cm}
\end{figure}

$~~~$ \vspace{3cm}

Fig. 6b

\bigskip
C. CIOFI DEGLI ATTI, E. PACE and G. SALME'
\newpage
\begin{figure}
\vspace{13cm}
%\special{picture prc7a}
\vspace {2cm}
\end{figure}

$~~~$ \vspace{3cm}

Fig. 7a

\bigskip
C. CIOFI DEGLI ATTI, E. PACE and G. SALME'
\newpage
\begin{figure}
\vspace{13cm}
%\special{picture prc7b}
\vspace {2cm}
\end{figure}

$~~~$ \vspace{3cm}

Fig. 7b

\bigskip
C. CIOFI DEGLI ATTI, E. PACE and G. SALME'
\newpage
\begin{figure}
\vspace{13cm}
%\special{picture prc7c}
\vspace {2cm}
\end{figure}

$~~~$ \vspace{3cm}

Fig. 7c

\bigskip
C. CIOFI DEGLI ATTI, E. PACE and G. SALME'
\newpage
\begin{figure}
\vspace{13cm}
%\special{picture prc8}
\vspace {2cm}
\end{figure}

$~~~$ \vspace{3cm}

Fig. 8

\bigskip
C. CIOFI DEGLI ATTI, E. PACE and G. SALME'
\newpage
\begin{figure}
\vspace{13cm}
\vspace {2cm}
\end{figure}

$~~~$ \vspace{3cm}

Fig. 9

\bigskip
C. CIOFI DEGLI ATTI, E. PACE and G. SALME'


\bigskip
\begin{thebibliography}{99}
\bibitem{1.}	T. W. Donnelly, A. S. Raskin, {\em Ann. Phys. (N.Y.)} {\bf 169},
247
(1986).
\bibitem{2.} a) C. E. Woodward
et al., {\em{Phys. Rev. Lett.}} {\bf{65}}, 698 (1990); b) C. E. Jones-Woodward
et
al., {\em Phys. Rev.} {\bf C 44}, R571 (1991); c) C. E. Jones-Woodward et
al., {\em Phys. Rev.} {\bf C 47}, 110 (1993).

\bibitem{3.} A. K. Thompson et
al., {\em Phys. Rev. Lett.} {\bf 68}, 2901 (1992); A. M. Bernstein,
{\em{Few-Body Systems Suppl.}} {\bf 6}, 485  (1992).
\bibitem{gao} H. Gao et al, {\em Phys.Rev.} {\bf{C}}, in press.
\bibitem{4.} B.
Blankleider and R.M. Woloshyn, {\em Phys. Rev.} {\bf C 29}, 538  (1984).
\bibitem{5.} C. Ciofi degli Atti, E. Pace and G.
Salm\`e, {\em Phys. Rev.} {\bf{C 46}}, R1591  (1992).
\bibitem{6.} R.W. Schultze and P.U. Sauer, {\em Phys. Rev.} {\bf C 48}, 38
(1993).
\bibitem{7.} C. Ciofi degli Atti, E. Pace and G.
Salm\`e, Proceedings of the VI Workshop on "Perspectives in Nuclear Physics at
Intermediate Energies", ICTP, May 1993, (World Scientific, Singapore,1994);
 C. Ciofi degli Atti, S. Scopetta, E. Pace and G.
Salm\`e, {\em Few-Body Systems Suppl.} {\bf 7}, 458 (1994).
\bibitem{8.} C. Ciofi degli Atti, E. Pace and G.
Salm\`e, {\em Phys. Lett.} {\bf B141}, 14 (1984).
\bibitem{10.} R.V. Reid, {\em Ann. Phys. (N.Y.)} {\bf 50}, 411 (1968).
\bibitem{11.} M. Lacombe et al., {\em Phys. Rev.} {\bf C21}, 861 (1980).
\bibitem{9.} C. Ciofi degli Atti, E. Pace and G.
Salm\`e, {\em Phys. Rev.} {\bf C 43}, 1155  (1991).
\bibitem{12.} M. Gari and W. Krumpelmann, {\em Z. Phys.} {\bf A322},  689
(1985);
{\em Phys. Lett.} {\bf B 173}, 10 (1986).
\bibitem{14.}	S. Blatnik and N. Zovko, {\em Acta Phys. Austriaca} {\bf 39}, 62
(1974).
\bibitem{15.}	G. Hoehler et al., {\em Nucl. Phys.} {\bf B 114},  505 (1976).
\bibitem{13.} S. Galster et al., {\em Nucl. Phys.} {\bf B32}, 221 (1971).
\bibitem{16.} D.A. Varshalovich, A.N. Moskalev, V.K. Khersonskii, "Quantum
Theory
of Angular momentum", World Scientific, Singapore, 1988.
\end{thebibliography}
 \end{document}